\documentclass{natureastroph}
\usepackage{graphicx}
\RequirePackage{times}
\RequirePackage{fullpage}
\RequirePackage{ifthen}

\title{ Ten percent polarized optical emission from GRB 090102}

\author{I.A. Steele$^{1}$, C.G. Mundell$^1$, R.J. Smith$^1$, S. Kobayashi$^1$, C. Guidorzi$^2$}
\begin{document}

\maketitle

\begin{affiliations}
\item  Astrophysics Research Institute, Liverpool John Moores University, CH41 1LD, U.K.
\item Physics Department, University of Ferrara, Via Saragat 1, 44122 Ferrara, Italy
\end{affiliations}

\begin{abstract}
{\bf The nature of the jets and the role of magnetic fields in gamma-ray bursts (GRB) remains unclear\cite{lyu06,zm04}. In a baryon-dominated jet only weak, tangled fields generated in situ through shocks would be present\cite{pir99}. In an alternative model, jets are threaded with large scale magnetic fields that originate at the central engine and which accelerate and collimate the jets\cite{ly09}. The way to distinguish between the models is to measure the degree of polarization in early-time emission, however previous claims of $\gamma$-ray polarization have been controversial\cite{cb03,rf03,wil05,McG07}.
Here we report that the early optical emission from GRB 090102 was polarized at the level of $P=10\pm1$\%, indicating the presence of large-scale fields originating in the expanding fireball. If the degree of polarization and its position angle were variable on timescales shorter than our 60-s exposure, then the peak polarization may have been larger than 10 per cent.
}
\end{abstract}

The standard GRB fireball model\cite{pir99} comprises an initial compact emitting region, expanding relativistically, in which internal shocks dissipate the bulk energy, converting kinetic to radiated energy, the so-called prompt emission.  As the shell of the relativistically expanding fireball collides with the surrounding circumburst medium, a forward shock is produced, which propagates outwards through the external medium and results in the long-lived afterglow whose emission is detectable from X-ray to optical, infrared and, in some cases, radio wavelengths.  Interaction of the relativistic fireball with the ambient medium also produces a short-lived reverse shock that propagates backwards through the expanding shell\cite{zk05,zkm03}.

Exploiting the ability of robotic optical telescopes to respond rapidly and automatically to the discovery of new GRBs, a  custom, fast-response, optical polarimeter\cite{ringo} (RINGO) was deployed on the 2.0 meter robotic Liverpool Telescope\cite{lt} (La Palma) with the goal of measuring the degree of polarization of optical emission from GRBs at early time. RINGO uses a rotating Polaroid to modulate the incoming beam, followed by corotating deviating optics that transfer each star image into a ring that is recorded on a CCD (Figure 1).  Any polarization signal present in the incoming light is mapped out around the ring in a $sin 2\theta$ pattern.  RINGO was first used in 2006, when it observed GRB~060418 at 203s after the gamma ray burst and coincident with the time of deceleration of the fireball.  At this time the reverse (assuming it was present) and forward shock components would have contributed equally to the observed light.  For GRB~060418 a 2$\sigma$ upper limit on optical polarization of P$<$8\% was measured in the combined light from the emitting regions\cite{mun07a}.  Until the burst reported here this was the only limit on early-time optical GRB polarization.

GRB~090102 was detected by the {\em Swift} satellite on 2 January 2009 at 02:55:45~UT, with a pulse of gamma rays lasting T$_{90}$=27 s and comprising four overlapping peaks starting 14-s before the trigger time\cite{man09}.  The automatic localization provided by the spacecraft was communicated to ground-based facilities, and a single 60-second RINGO exposure was obtained starting 160.8 seconds after the trigger time.  Simultaneously with our polarization observation of GRB~090102, a number of automated photometric followups were also performed by other facilities\cite{man09,kl09,cov09}. 
The optical light curve, beginning at 40-s postburst, can be fitted by a broken power law whose flux density ($F$) decays as a function of time ($t$) such that $F\propto t^{-\alpha}$ with gradient $\alpha=1.50\pm0.06$ that then flattens to $\alpha=0.97\pm0.03$ after approximately 1000~s\cite{gen09}.  In contrast, the X-ray light curve, begun at 396~s after the GRB due to observing constraints, shows a steady decay consistent with a single power law with slope $\alpha$=1.36$\pm$0.01 and no evidence of flares or breaks up to t$>$7$\times$10$^5$~s post-burst\cite{man09}. The absence of any additional emission components from late-time central engine activity superimposed on the afterglow light curve allows a straightforward interpretation of the light curves in the  context of current GRB models.  
The steep-shallow decay of optical emission from GRB~090102 is characteristic of an afterglow whose early-time light is dominated by  fading radiation generated in the reverse shock\cite{zkm03,gom08}.

Figure 1 shows the RINGO exposure obtained on the night of 2009 January 2.
The afterglow of GRB 090102 is clearly visible, as are six brighter foreground objects. Detection of these objects allowed us to perform additional checks on the instrumental calibration at the time of the GRB.
In addition by observing the same region of sky at later dates after the GRB had faded (2009 January 28, 2009 April 18 and 2009 May 19) the stability of RINGO was also verified. The measured optical (4600 - 7200{\AA}) polarization of GRB 090102 is $P=10.2\pm1.3$\%, in contrast to the foreground objects 
which show $P\sim1-4$\% (Figure 2). A simple Monte Carlo analysis (Figure 3) was performed to 
estimate the significance of the polarization measurements.   This showed that the rank of our GRB measurement amongst a distribution of randomly reordered GRB trace data was 9,988/10,000.

In interpreting our measurement first we consider whether such a polarization could be produced via the production of magnetic instabilities in the shock front  (Figure 4(c)).  A very optimistic estimate of the coherence length can be made by assuming it grows at about the speed of light in the local fluid frame after the field is generated at the shock front - in this situation polarized radiation would come from a number of independent ordered magnetic field patches.  A measured polarization of 10\% is at the very uppermost bound for such a model\cite{gw99} and therefore seems unlikely. As an alternative to the ``patch'' model, we have also considered the case where the observer's line of sight is close to the jet edge\cite{gru99} (Figure 4(b)).  In this case since the magnetic fields parallel and perpendicular to the shock front could have significantly different averaged strengths\cite{ml99} a polarization signal can also be produced.  However applying this model to GRB~090102 we would have expected a steepening of the light curve (a ``jet-break'') just after the time of our polarization measurement rather than the observed flattening.  Similarly there is no evidence of a jet break in the X-ray light curve up to late times.  Furthermore, our detection of 10\% is much higher than the reported polarization signal of a few $\%$ associated with a jet break in late time afterglow of other events\cite{cov99, wij99}.  We also rule out an Inverse Compton origin for the optical polarization - a mechanism suggested to explain earlier $\gamma$-ray polarization measurements\cite{laz04} - in which lower energy photons are scattered to higher energies by colliding with electrons in the relativistic flow. If Inverse Compton emission is present, it is more likely to contribute to the high-energy X-ray and $\gamma$-ray bands than the optical band and again requires the observer's line of sight to be close to the edge of the jet (Figure 4(b)) to produce significant polarization which as we have already discussed is not the case for GRB~090102.

It therefore seems apparent that in the case of GRB~090102 the high polarization signal requires the presence of large-scale ordered magnetic fields in the relativistic outflow (Figure 4(a)) . As the measurement was obtained while the reverse-shock emission was dominant in GRB~090102, the detection of significant polarization provides the first direct evidence that such magnetic fields are present when significant reverse shock emission is produced.
Magnetization of the outflow can be expressed as a ratio of magnetic to kinetic energy flux, $\sigma$.  The degree of magnetization cannot
be sufficient for the jet to be completely Poynting flux dominated ($\sigma>1$)
since this would be expected to suppress a reverse shock\cite{mga09}.
We can therefore reconcile the
detection of polarization in GRB~090102 and our previous non-detection in GRB~060418 
in a unified manner if GRB jets have magnetization 
of $\sigma \sim 1$. In the GRB~060418 case, the jet would have had slightly higher magnetization than unity, resulting in the suppression of a reverse shock, while GRB~090102
would have $\sigma$ slightly smaller than unity, which is optimal to 
produce bright reverse shock emission.  Of course due to the small sample (only two 
objects), we can not rule out a possibility that each GRB jet had very different magnetization.

Finally we note that a high degree of polarization is also predicted for the prompt $\gamma$-ray emission in the presence of large-scale ordered magnetic fields\cite{gran03, fan09}.  Recent claims of rapidly ($\sim10$ s) variable $\gamma$-ray polarization from less than 4\% to 43\% ($\pm$25\%) in the prompt emission of GRB~041219A\cite{gotz09} lend further support to models with magnetized outflows and offer the possibility that the peak optical polarization from GRB~0901012
could have been even higher than that measured in our 60 second exposure.

\noindent
{\bf Acknowledgements}
The Liverpool Telescope is operated on the island of La Palma by Liverpool John Moores University in the Spanish Observatorio del Roque de los Muchachos of the Instituto de Astrofisica de Canarias with financial support from the UK Science and Technology Facilities Council. 
CGM acknowledges financial support from the Royal Society and Research Councils U.K.  We thank Dr Jon Marchant for preparing Figure 4 for us.

\noindent
{\bf Author Contributions}
IAS:   Instrument design and build.  Data reduction and analysis.  \\
CGM: Instrument science case, scientific interpretation.\\
RJS:  Data reduction and analysis.\\
SK:  Theoretical interpretation.\\
CG:  Coordination of observations and identification of afterglow.

\noindent
{\bf Competing Interests} The authors declare that they have no competing financial interests.

\noindent
{\bf Correspondence} Correspondence and requests for materials
 should be addressed to IAS.

(email: ias@astro.livjm.ac.uk)
\newpage

\begin{figure}
\centering\includegraphics[scale=0.7]{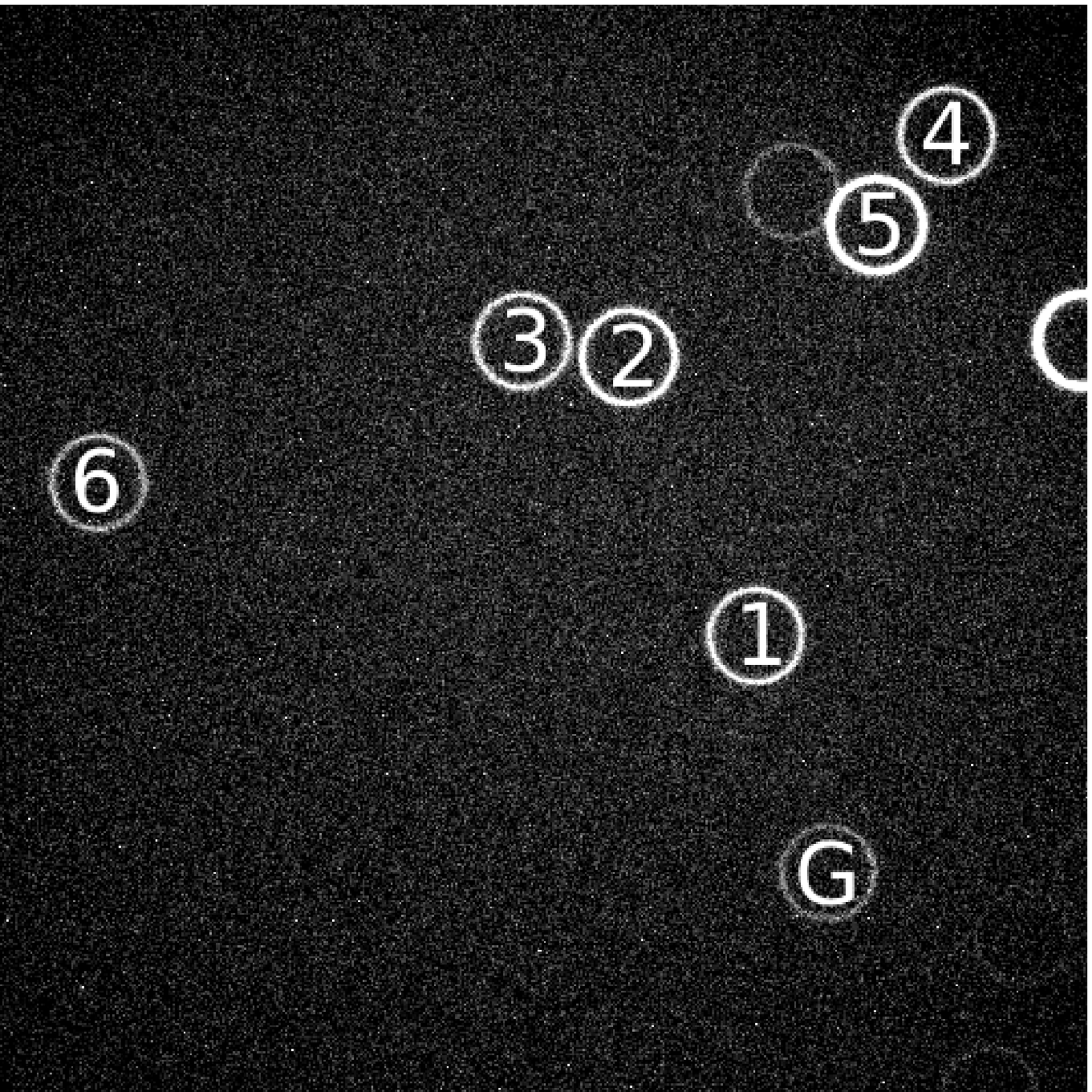}
\\
\noindent
{\bf Figure 1.
RINGO observation of the field of GRB~090102 observed 2009 Jan 2.}  The field of view is 4.6 x 4.6 arcmin.  The data have been dark subtracted and flat-fielded using standard astronomical algorithms.  The afterglow of GRB 090102 is labeled G along with six foreground sources (labeled 1--6).  Foreground source 5 is contaminated by an overlapping faint source, and so was not used in further analysis.  We followed our standard RINGO reduction procedure in which flux traces for all objects on all nights were extracted within annuli with inner (8 arcsec) and outer (14 arcsec) radii sufficient to ensure that seeing variations do not influence the extracted fluxes.  The traces were then sky subtracted by the normalized flux inside the inner trace radius and divided through by an average of the traces from routinely obtained zero-polarization standards\cite{sch92} to remove the known 2.7\% instrumental polarization.  The resulting flux traces for a sample of objects and the GRB are presented in Figure 2.
\end{figure}

\newpage

\begin{figure}
\centering\includegraphics[scale=0.6,angle=270]{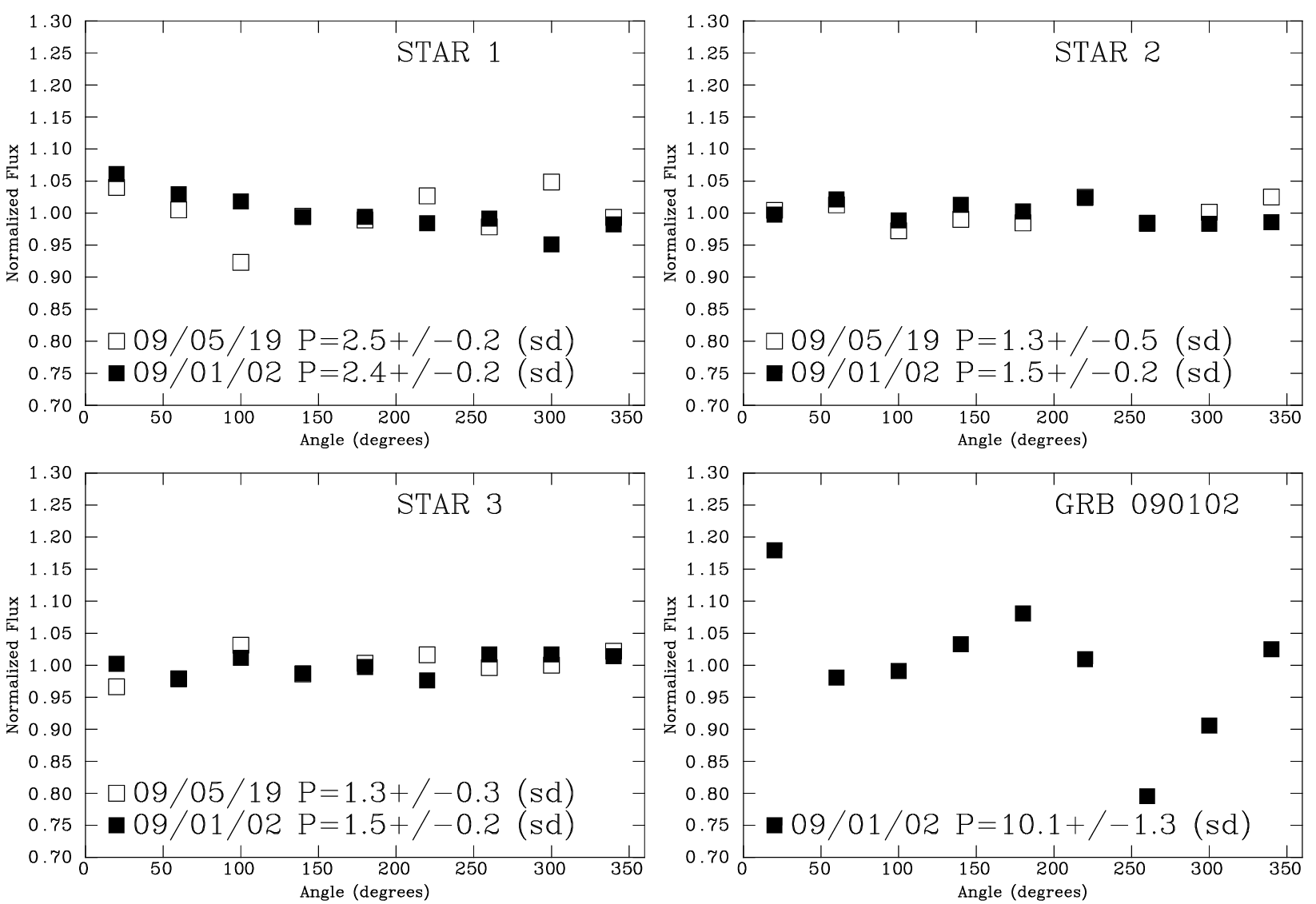}
\\
\vspace{5mm}
\noindent
{\bf Figure 2.
RINGO data for GRB 090102 and calibration sources.}
Example flux traces around the rings of three of the foreground objects (1-3) and GRB~090102 showing a clear $\sin 2\theta$ signal for the GRB are shown.   For the foreground objects traces are presented taken simultaneous with GRB 090102 (filled symbols) and on the night of 2009 May 19 (unfilled symbols). Analysis\cite{cd02} of different subsets\cite{mun07a} of the data in the GRB trace allows a mean polarization and standard deviation to be measured, giving a value of $10.1\pm1.3$\% for GRB 090102.  Objects 2 and 3 have low polarization ($\leq1.5$\%) in both exposures and set limits on uncorrected instrumental polarization effects.  Object 1 is detected as weakly polarized ($2.5$\%) in both measurements, demonstrating the stability of the instrumental setup (the instrument reference position angle varies due to the telescope altitude-azimuth mount between the two epochs, so the traces are not in phase, however the amplitude of variation and hence derived polarization is similar).  Objects 4 and 6 (not plotted) show similar stable weak polarization between different epochs of $\sim3$\% and $\sim4$\% respectively.
\end{figure}
\newpage

\begin{figure}
\centering\includegraphics[scale=0.7,angle=270]{fig3.ps}
\\
\vspace{5mm}
\noindent
{\bf Figure 3.
Monte Carlo simulation using GRB090102 data.}
The distribution of measured polarizations derived from randomly reordered GRB trace data demonstrates the significance of the actual result.  The measured value for the GRB (10.1\%) is shown with an arrow, and can be seen to be highly significant (rank = 9,988/10,000).  Similar analyses for the foreground objects in the frames confirms that objects 2 (rank 780/1,000) and 3 (rank 540/1,000) have no detectable polarization
at the level of 1.5\% and that objects 1 (2.5\% - rank 969/1,000) , 4 (3.3\% - rank 927/1,000) and 6 (4.1\% - rank 913/1,000) have measured polarizations in line with the expected values for stars within our galaxy\cite{sch92}.
\end{figure}
\newpage

\begin{figure}
\centering\includegraphics[scale=0.7,angle=0]{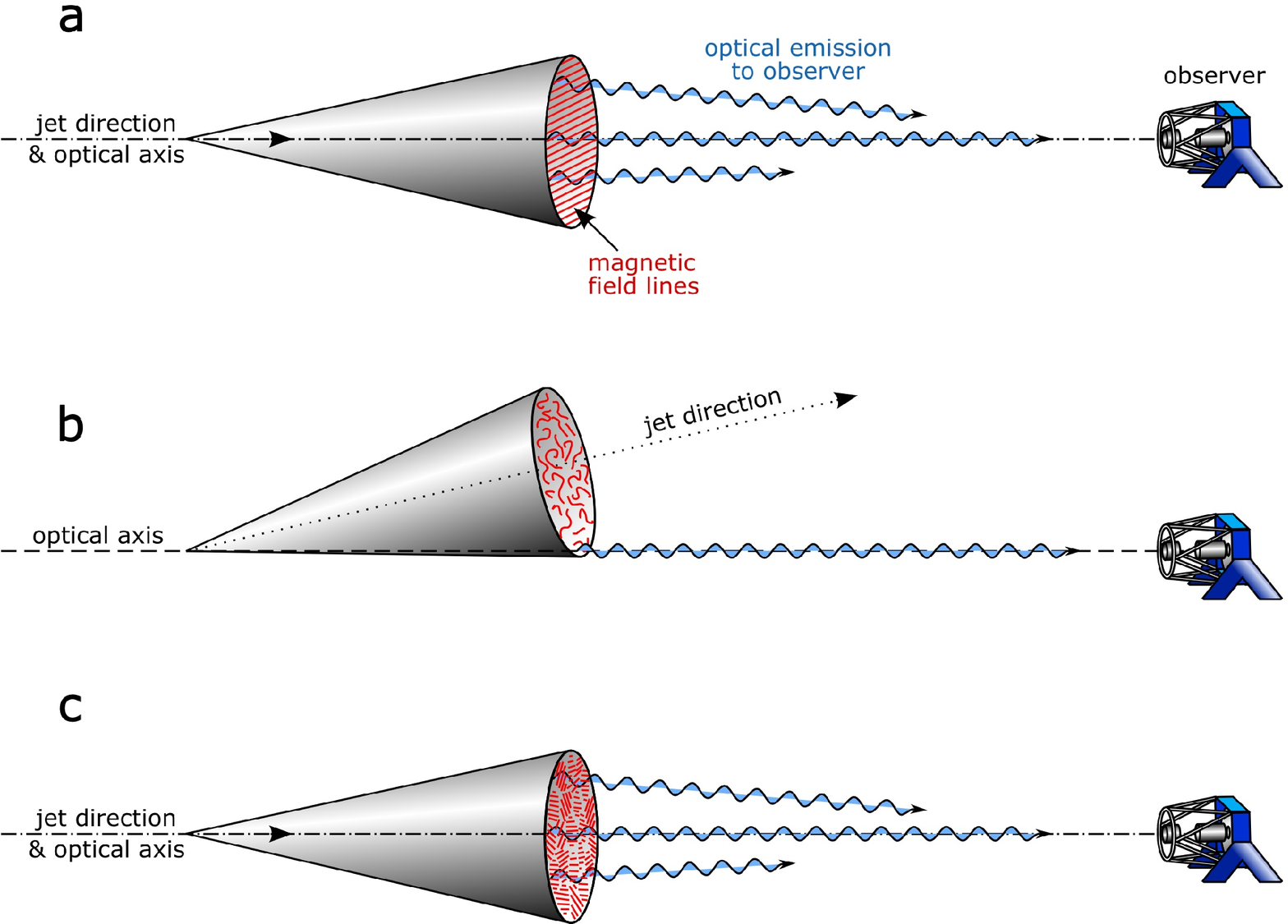}
\\
\vspace{5mm}
\noindent
{\bf Figure 4. 
Competing models of GRB magnetic field structure.}  The
schematic shows three representations of a GRB outflow in the context of the standard fireball model for a variety of magnetic field structures and different orientations to the observer's line of sight (optical axis). A large degree of polarization is predicted when the ejected material is threaded with a large-scale ordered magnetic field as shown in (a) and is the favoured model to explain the measured polarization in GRB~090102. Alternatively, if no ordered magnetic field is present and, instead a tangled magnetic field is produced in the shock front, the detected light will be polarized only if the observer's line of sight is close to the jet edge (b). In this case, however, the predicted {\em steepening} of the light curve that is expected when observing an off-axis jet is inconsistent with the {\em flattening} exhibited in the light curve of GRB~090102.    A compromise is shown in (c) in which the shock front contains a number of independent patches of locally-ordered magnetic fields; a measured polarization of 10\% is at the very uppermost bound for such a model.
\end{figure}
\clearpage

\end{document}